# Electronic and Magnetic Properties of Small Fullerene Carbon Nanobuds: A DFT Study

Amrish Sharma[1], Sandeep Kaur[1], Hitesh Sharma[2], Isha Mudahar[3*]
[1]Department of Physics, Punjabi University, Patiala.
[2]Department of Physics, IKG Punjab Technical University, Kapurthala.
[3]Department of Basic and Applied Sciences, Punjabi University, Patiala.
*dr.ishamudahar@gmail.com

**Abstract**: The electronic and magnetic properties of carbon nanobuds have been investigated using density functional theory. The carbon nanobuds are formed by attaching smaller fullerenes ($C_{20}$, $C_{28}$, $C_{36}$ and $C_{40}$) of variable size with (5,5) ACNT and (5,0) ZCNT. Fullerenes interact strongly with CNT surface having binding energies within the range -0.93eV to -4.06eV. The C-C bond lengths near the attachment region increase from the original C-C bond lengths. The relative stabilities of the nanobuds are closely related to C-C bond lengths and bond angles in cycloaddition reaction. Nanobuds formed by bond cycloaddition are energetically most favorable amongst all cycloadditions. The electronic and magnetic properties of nanobuds depend strongly on electronic properties of its building blocks. The attachment of $C_{20}$ and $C_{40}$ on CNTs open up the HOMO-LUMO gaps of nanobuds whereas $C_{28}$ and $C_{36}$ results in addition of impurity states near the Fermi level. The total magnetic moment of nanobuds vary from $0.28\mu_B$ to $4.00\mu_B$ which depend on the nature of bonding between fullerene and CNTs. The results outline the potential of nanobuds as hybrid carbon nanostructures and how their properties can be tuned with the size and type of fullerene attached.

## Introduction

Carbon nanobuds (CNBs) are new class of carbon nanostructures, which has attracted a lot of interest in recent years due to possibility of combining functionalities of its building blocks [1-2]. The CNBs can be formed by combining nanostructures such as fullerenes, carbon nanotubes (CNTs) and graphene. Nanobud structure which can act as molecular anchors through fullerenes can modify electronic, magnetic and mechanical properties of CNTs [2, 7-9].

CNBs have been synthesized during a process in which fullerenes ($C_{60}$) were formed on iron-catalyst particles together with CNTs during CO disproportionation [2]. The Raman spectroscopy of CNBs have confirmed the presence of fullerenes, CNTs, covalent bonding of fullerene with CNTs and predicted their lower thermal stability than pristine single walled carbon nanotube (SWCNT) [3,4].

In addition to pristine CNBs, functionalized CNBs have also been synthesized using nucleophilic addition to CNB fullerenes where the most stable structure is formed when additions are at ortho position w.r.t a six membered ring of fullerene cage [5]. Field emission characteristics investigation of CNBs has revealed the current density of order of 10 A/cm$^2$ predicting them for applications in high-current vacuum electronic devices [6]. The studies further predict high possibility of CNB formation with smaller fullerenes [2]. Therefore, combinations of chirality and size of CNTs with fullerenes of variable surface curvature offers a new class of carbon based hybrid structures which can be synthesized with tunable electronic properties.

CNTs have been investigated extensively in the literature which shows that they can be synthesized in the wide range of diameter. The smallest CNTs synthesized is with (3,0) and (3,3) chirality [10-12] and exhibit either metallic or semiconducting behavior depending on its helicity and chirality [12-14]. Similarly, fullerenes have also been synthesized from small diameter $C_{20}$ to giant $C_{266}$ [15-20]. The fullerenes smaller than $C_{60}$ have been reported in the gas phase experiments due to their higher chemical reactivity [21-33]. $C_{36}$ has been predicted as one of the magic number small fullerene detected by mass spectroscopy. It possesses fifteen isomers and amongst them $D_{6h}$ symmetry is most stable [34-36]. Further, $C_{36}$ is highly reactive and possess a strong tendency to form intermolecular covalent bonds [37-38]. Hydrogenated derivative of $C_{40}$ has been synthesized and theoretical investigation have shown forty symmetric isomers, out of which $D_2$ and $T_d$ symmetry are the most stable [35,39]

To best of our knowledge, smaller fullerene based CNBs have not been studied systematically except for $C_{20}$-CNBs [40]. It is therefore of significant interest to investigate the change in the structural, electronic and magnetic properties of CNBs. The change in their properties with respect to their building blocks is of fundamental importance for tuning of their properties. In this paper, we have investigated the possibility of CNBs formation by combining CNTs with small fullerene cages like $C_{20}$, $C_{28}$, $C_{36}$ and $C_{40}$ using DFT based ab-initio calculations.

**Computational Details**

We have used the Spanish Initiative for Electronic Simulation with Thousands of Atoms (SIESTA) computational code [41-43] which is based on numerical atomic orbital density functional theory [44]. The calculations are carried out by using Generalized Gradient Approximation (GGA) that implements Perdew, Burke and Ernzerhof (PBE) exchange – correlation [45]. Core electrons are replaced by non-relativistic, norm-conserving pseudopotential generated by improved Troullier Martins Scheme [46]. The valence electrons are described using linear combination of numerical pseudo atomic orbitals of Sankey-Niklewski type [47] but generalized for multiple-$\zeta$ and polarization function. Along the tube axis, we have taken six unit cells for armchair SWCNT and five unit cells for zigzag SWCNT. Split valence double-$\zeta$ polarized (DZP) basis set with an energy cutoff of ≈ 250 - 300 Ry has been used. The structures are obtained by minimization of total energy using Hellmann-Feynman forces, including Pulay like corrections, until the residual forces acting on each atom were smaller than 0.03 eV/Å.

Test calculations were performed on small fullerenes ($C_n$, n ≤ 40) and CNTs which are tabulated in Table 1. Our results are in good agreement with the previous studies [48-50]. Since our group has already study the carbon based systems so the parameters have been checked [51-55].

**Table 1. Average diameter ($D_{av}$), Average Bond Length ($A_vB_L$) and HOMO-LUMO gaps ($E_{gap}$).**

| Fullerenes/CNTs | $D_{av}$ (Å) | $A_vB_L$ (Å) | $E_{gap}$ (eV) |
|---|---|---|---|
| $C_{20}$ | 4.14 | 1.48 | 0.75 |
| $C_{28}$ | 4.83 | 1.49 | 0.33 |
| $C_{36}$ | 5.60 | 1.46 | 0.43 |
| (5,5) | 6.78 | 1.42 | Metallic |
| (5,0) | 3.92 | 1.42 | 0.27 |

To investigate the interaction strength, the binding energy ($E_b$) of CNBs is computed as

$$[E_b = E_{CNB} - E_{CNT} - E_{fullerene}]$$

where $E_{CNB}$, $E_{CNT}$ and $E_{fullerene}$ denotes the total energies of CNB, CNT and fullerene respectively.

### **Results and Discussion**

$C_{20}$ fullerene consists of pentagonal rings of carbon atom whereas all other fullerene cages have both pentagonal and hexagonal faces. Three types of C-C bonds exist in the fullerenes: C-C bond between pentagonal-pentagonal ($A_5$-$A_5$) face, C-C bond between hexagonal-pentagonal face ($A_6$-$A_5$) and C-C bond between hexagonal-hexagonal face ($A_6$-$A_6$) (Fig.1 (a-c)). The fullerenes are bonded to sidewall of both armchair CNT (ACNT) and zigzag CNT (ZCNT). We have optimized (5,5) ACNT and (5,0) ZCNT of finite length 14Å and 24Å respectively. The C-C bonds in ACNT or ZCNT is shown in perpendicular direction (marked as S) or parallel direction (marked as P) and tilted direction (marked as T) along the tube axis (Fig. 2(a-b)). The edge atoms of CNTs are passivated from both sides with H atom so as to reduce the edge effects.

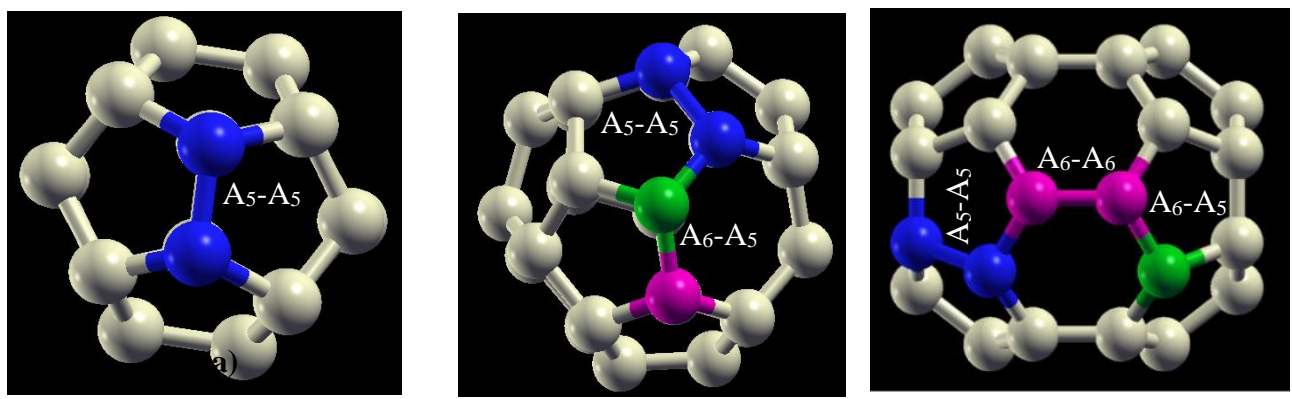

**Fig. 1 (in colour); Ball and Stick model for (a) $C_{20}$ (b) $C_{28}$ (c) $C_{36}$, same for $C_{40}$; blue balls specify the C-C bond connecting pentagon-pentagon ($A_5$-$A_5$) rings, pink and green balls specify the C-C bond connecting hexagon-pentagon ($A_6$-$A_5$) ring, Pink balls specify C-C bond connecting hexagon-hexagon ring ($A_6$-$A_6$).**

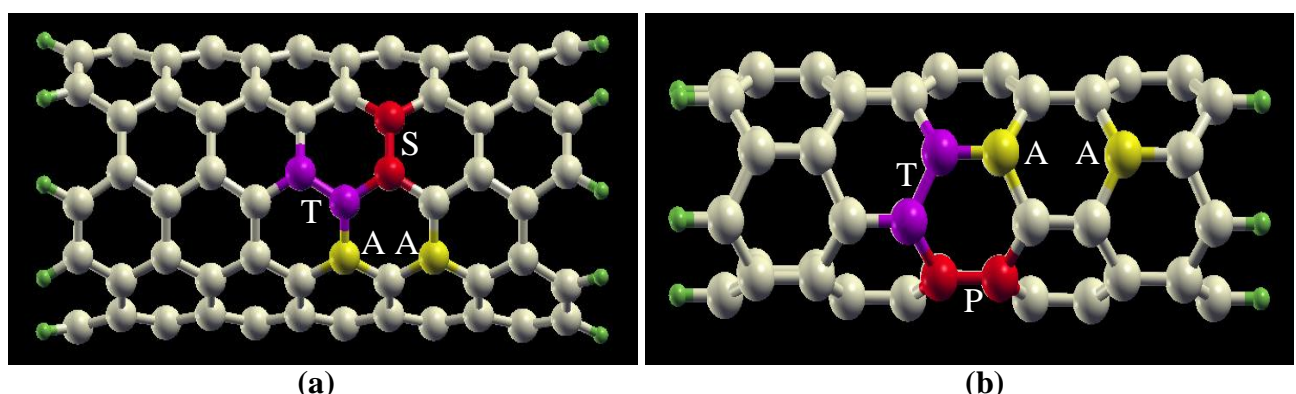

**Fig. 2 (in colour); Ball and Stick model for (a) ACNT (5,5) (b) ZCNT (5,0); violet balls specify C-C bond in tilted position, maroon balls specifies C-C bond in perpendicular or parallel position, golden balls specify contrapuntal carbon atom marked as AA.**

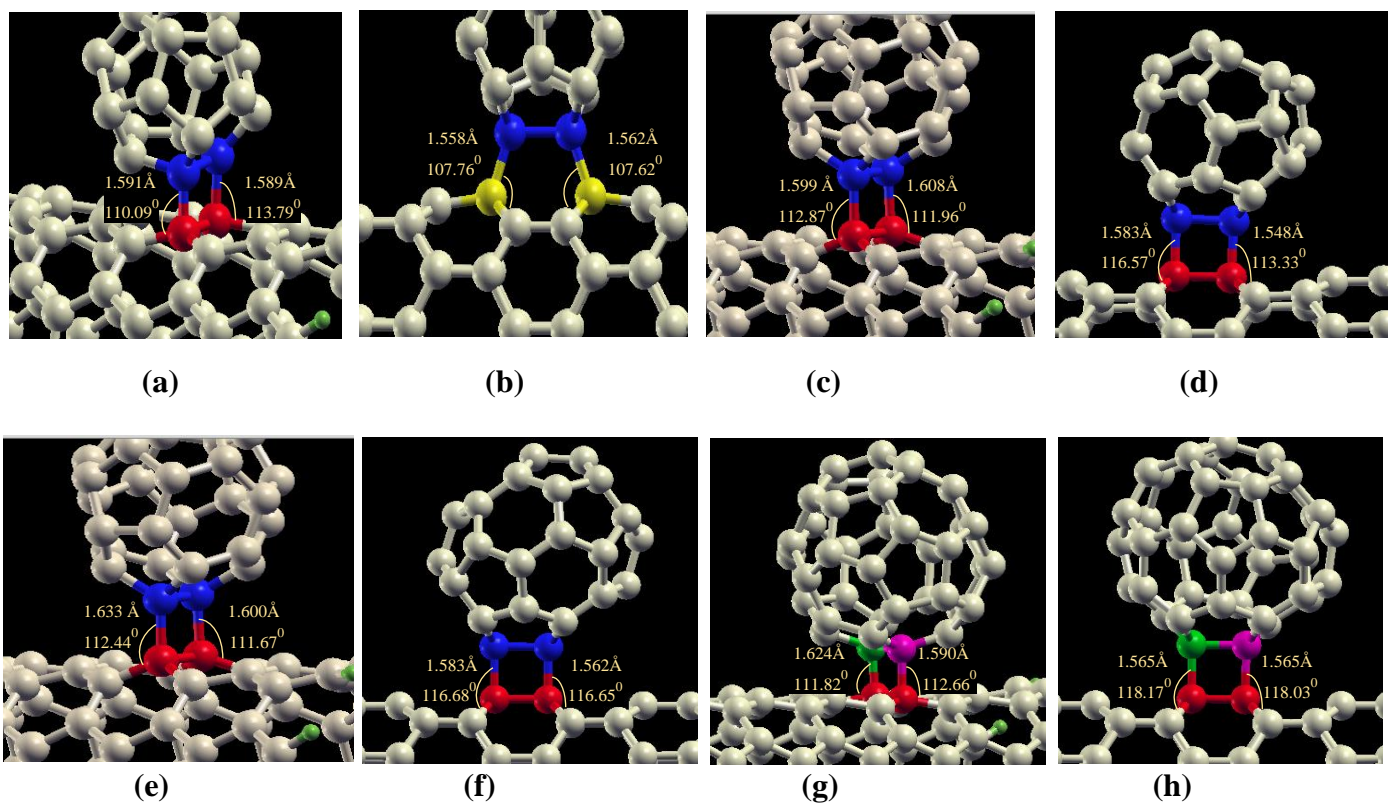

(a)　　　(b)　　　(c)　　　(d)

(e)　　　(f)　　　(g)　　　(h)

**Fig. 3 (in colour); Most Stable CNB configurations with bond lengths and bond angles (a) $C_{20}$-(5,5) (b) $C_{20}$-(5,0) (c) $C_{28}$-(5,5) (d) $C_{28}$-(5,0) (e) $C_{36}$-(5,5) (f) $C_{36}$-(5,0) (g) $C_{40}$-(5,5) (h) $C_{40}$-(5,0).**

**$C_{20}$-CNB**

$C_{20}$ fullerene having $I_h$ symmetry is bonded to ACNT or ZCNT through different cycloaddition configurations: (a) the carbon atom of $C_{20}$ is bonded to the carbon atom of ACNT or ZCNT to form [1+1] atom cycloaddition (b) the C-C bond of $A_5$-$A_5$ face in $C_{20}$ is bonded to C-C bond of ACNT or ZCNT in perpendicular or parallel and tilted configuration (c) the C-C bond of $A_5$-$A_5$ face is bonded to contrapuntal atoms (Fig. 2(a-b)) of hexagonal ring of ACNT or ZCNT and such type of cycloadditions are called [2+2] bond cycloadditions. In all configurations of $C_{20}$-CNBs, the initial distance between fullerene and CNT is considered to be 1.45Å. All configurations are optimized and there is induction of local distortion at the junction of CNTs due to covalent bonding between $C_{20}$ and CNT and carbon atoms of CNTs near $C_{20}$ were pulled outward from original wall surface. The average connecting C-C bond length ($A_vB_L$) between $C_{20}$ and CNT increases from 1.45Å to 1.58Å and this change is due to the change in hybridization from $sp^2$ to $sp^3$.

The [1+1] covalent bond between fullerene and CNT in atom cycloaddition is unstable and can spontaneously break so at least two carbon atoms of both are required to be bonded together to form a stable complex [56]. $C_{20}$-CNBs formed through atom cycloaddition have binding energies -1.55eV and -2.20eV for ACNT or ZCNT base respectively, which shows their stability might be due to high chemical reactivity of $C_{20}$. Table 2 shows the calculated binding enery ($E_b$), $A_vB_L$ and average bond angle ($\underline{A}_vB_A$) for the most stable configuration of CNB structures. Negative values of $E_b$ show a strong interaction between fullerene cage and CNT which lead to higher stability of the complex. All $C_{20}$-CNBs configurations have negative $E_b$ and amongst them [2+2] $A_5$-$A_5$/S configuration (Fig.3a) where carbon atoms of pentagonal-pentagonal face in $C_{20}$ are bonded to C-C bond of ACNT perpendicular to tube axis in the form of quadrilateral is most stable. However, for ZCNT base [2+2] $A_5$–$A_5$/AA configuration (Fig.3b) in which the carbon atoms of pentagonal-pentagonal face in $C_{20}$ are bonded to contrapuntal atoms in carbon ring of CNT to form a stereo hexagon is the most stable one which is in good agreement with the previous study [40]. The stability of different configurations of $C_{20}$-CNBs can be examined with the deviation in $\underline{A}_vB_A$ of $sp^3$ bonds at the junction. The ideal $B_A$ for $sp^3$ hybridization is $109.5^0$ and for [2+2] $A_5$-$A_5$/S in ACNT and [2+2] $A_5$–$A_5$/AA in ZCNT, the closest $\underline{A}_vB_A$ is near to this ideal value. In general, it can be predicted that $C_{20}$-CNBs formed through bond cycloaddition have high stability than formed through atom cycloaddition. This might be due to a weak driving force as a result of formation of four C atom ring at the junction of $C_{20}$-CNB for [2+2] bond cycloaddition.

## $C_{28}$-CNB

$C_{28}$ fullerene with $T_d$ symmetry was considered for the formation of $C_{28}$-CNBs. In addition to all $C_{20}$-CNB configurations, except the one where carbon atoms of $C_{20}$ are connected to contrapuntal atoms of CNT, two more configurations are possible for $C_{28}$-CNBs i.e. (a) the C-C bond of $A_6$-$A_5$ face in $C_{28}$ are bonded to the C-C bond of ACNT or ZCNT in perpendicular or parallel and tilted configuration, (b) the carbon atoms in hexagonal ring of fullerene are bonded to the carbon atoms in hexagonal ring of ACNT or ZCNT through [6+6] ring cycloaddition.

Similar to $C_{20}$-CNB, $C_{28}$ cage also has a tendency to move away from CNT and $A_vB_L$ of the most stable $C_{28}$-CNBs is nearly same as $C_{20}$-CNBs. The $E_b$ of all $C_{28}$-CNB configurations is negative, which point towards the favorability of their formation. It was previously reported that [6+6] ring cycloaddition for $C_{60}$-CNB is unfavorable with positive $E_b$ [56] but in $C_{28}$-CNBs we find ring cycloaddition is also favorable with $E_b$ values -1.00eV and -0.70eV for ACNT and ZCNT respectively. Further, with ACNT base the most stable configuration is similar as of $C_{20}$-CNB (Fig.3c), whereas for ZCNT base [2+2] $A_5$–$A_5$/P configuration (Fig.3d) where carbon atoms of $A_5$-$A_5$ face in fullerene are bonded to C-C bond parallel to tube axis in forming a square is the most stable.

## C$_{36}$-CNB

C$_{36}$ in D$_{6h}$ symmetry was considered for the formation of C$_{36}$-CNBs. In addition to C$_{28}$-CNB configurations, we considered another [2+2] configuration where the C-C atom of A$_6$-A$_6$ face in fullerene is bonded to C-C atom of ACNT or ZCNT in perpendicular or parallel and tilted configuration. C$_{36}$-CNBs follow same pattern for A$_V$B$_L$ as C$_{20}$-CNBs. The positive value of E$_b$ for C$_{36}$-CNB with ACNT or ZCNT base in [6+6] ring cycloaddition (2.99eV, 1.36eV) implies that C$_{36}$-CNB formed by this cycloaddition are unfavorable. The [1+1] atom cycloaddition and [2+2] bond cycloaddition configurations show negative E$_b$ and the most stable configurations with ACNT or ZCNT base is similar (Fig.3(e-f)) as in C$_{28}$-CNB.

## C$_{40}$-CNB

C$_{40}$-CNBs are formed using D$_2$ symmetry of C$_{40}$ fullerene and all the configurations considered for the formation of C$_{40}$-CNB are similar to that of C$_{36}$-CNB. C$_{40}$ also has a tendency to move away from CNT wall and the A$_V$B$_L$ for ACNT or ZCNT base are 1.60Å and 1.56Å respectively. The E$_b$ values for all C$_{40}$-CNB configurations are negative except for [6+6] ring cycloaddition having values 1.90eV and 1.60eV for both ACNT or ZCNT bases. C$_{40}$ forms most stable nanobuds when it is connected in [2+2] A$_6$-A$_5$/S (Fig.3g) configuration with ACNT base and [2+2] A$_6$-A$_5$/P (Fig.3h) configuration with ZCNT base. In these configurations, the carbon atoms of A$_6$-A$_5$ face of C$_{40}$ are bonded to C-C bond perpendicular or parallel to tube axis forming a square. The E$_b$ values go on increasing from C$_{20}$ to C$_{40}$ making nanobuds relatively less stable (Table 2). The stabilities of all nanobuds in their most stable configurations can be seen from A$_V$B$_L$ between fullerenes and CNTs as well as their A$_V$B$_A$ in sp$^3$ hybridization. Shorter the B$_L$ stronger will be the binding and closer the value of hybridized B$_A$ to ideal value more will be the stability of nanobuds.

In order to confirm the most stable configuration of the nanobuds, the energy relative to energy of most stable configuration for all cycloadditions is calculated by changing the distance between fullerenes and CNTs. The graph is plotted between energy and distance between these two (Fig.4) which shows a typical landslide with the minimum laying at the distance of 1.60Å for ACNT base and 1.56Å for ZCNT base.

**Table 2. Binding Energy ($E_b$), Average Bond Length ($A_vB_L$), Average Bond Angle ($A_vB_A$) of the most stable CNBs.**

| Carbon Nanobuds $C_n$-CNBs | Type of CNT | Most stable configuration of CNBs | $E_b$(eV) | $A_vB_L$(Å) | $A_vB_A$(Degree) |
|---|---|---|---|---|---|
| $C_{20}$-CNBs | Armchair | [2+2] $A_5$-$A_5$/S | -2.56 | 1.59 | 111.94 |
| | Zigzag | [2+2] $A_5$-$A_5$/AA | -4.06 | 1.56 | 107.69 |
| $C_{28}$-CNBs | Armchair | [2+2] $A_5$-$A_5$/S | -1.84 | 1.60 | 112.41 |
| | Zigzag | [2+2] $A_5$-$A_5$/P | -3.20 | 1.56 | 114.95 |
| $C_{36}$-CNBs | Armchair | [2+2] $A_5$-$A_5$/S | -0.93 | 1.61 | 112.06 |
| | Zigzag | [2+2] $A_5$-$A_5$/P | -2.55 | 1.57 | 116.67 |
| $C_{40}$-CNBs | Armchair | [2+2] $A_6$-$A_5$/S | -0.97 | 1.60 | 112.24 |
| | Zigzag | [2+2] $A_6$-$A_5$/P | -2.61 | 1.56 | 118.10 |

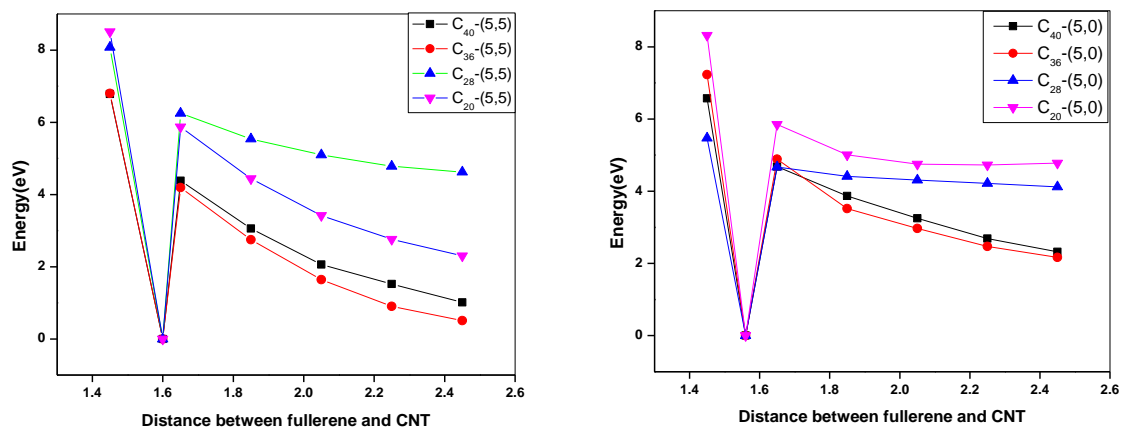

**Fig.4 (in colour); Energy relative to energy of most stable configuration by varying the distance between fullerenes and CNTs (a) fullerenes-(5,5) ACNT (b) fullerenes-(5,0) ZCNT. The distance between fullerene and CNT is in Å.**

**Electronic and Magnetic Properties**

The electronic and magnetic properties of small fullerene CNBs have been investigated using spin polarized density functional theory. The total magnetic moment (TMM), highest occupied molecular orbital (HOMO) and lowest unoccupied molecular orbital (LUMO) gaps (with Perdew, Burke and Ernzerhof (PBE) and Lee, Young and Parr (BLYP) functional) and net charge transfer (q) have been calculated (Tables 3, 4 and 5). The DFT based calculations done here are used to describe the ground state properties of the magnetic system accurately. However, at finite temperature, it becomes difficult to calculate these properties due to limitations of functionals. Therefore, the magnetic behavior of nanobuds calculated at 0K, which suggest small gap between ground and lowest excited state of various spin states may behave differently at finite temperature. Further, the literature survey shows that description of electrons at finite temperature in DFT is still an unaddressed problem at the fundamental level. Accurate description of finite-temperature contributions in DFT requires accurate description of approximate functionals. The calculations of magnetic phase transitions is difficult in current spin dependent DFT due to low-lying collective excitations which requires a non-collinear description of spin. Hence, a finite-temperature version of spin DFT involving only spin-up and spin-down densities have not been able to predict properties like critical temperature [57]. Limited efforts have been made to include finite temperature effects in DFT for all finite temperature Kohn Sham treatments by constructing accurate approximations for ground-state DFT which are generalized to finite temperature, including the adiabatic connection formula [58].

$C_{20}$ and ZCNT are non-magnetic in isolated form, whereas ACNT has TMM of $3.8\mu_B$. When $C_{20}$ connects either with ACNT or ZCNT, the resultant nanobud is found to be non-magnetic in its most stable configuration. However, when ACNT or ZCNT is connected with $C_{20}$ through [1+1] atom cycloaddition mode, the nanobud has TMM of $0.18\mu_B$ and $0.28\mu_B$ respectively. For $C_{20}$-(5,5) in [1+1] mode, the local MM show that TMM of $C_{20}$ cage is $0.54\mu_B$ which is reduced by antiferromagnetic interactions between C atoms in ACNT. The major contribution to TMM for $C_{20}$-(5,5) in [1+1] mode is given by first, second and third nearest neighbors (NN) from connecting bond which contribute 16%, 7% and 10% of TMM respectively. The $C_{20}$-(5,0) in [1+1] mode show major contribution to TMM comes from first NN~49%, whereas connecting bond atoms don't show any magnetic contribution as the MM is localized away from them. In order to visualize the magnetic ordering, the electron spin distribution of CNBs have been plotted in figure 5. Figure 5(a) show symmetrical spin charge density on both subunits w.r.t [2+2] cycloaddition which result in nonmagnetic [2+2] $A_5$-$A_5$/S configuration of $C_{20}$-(5,5). The HOMO-LUMO gap ($E_{gap}$) for spin up (↑) and spin down (↓) electrons have equal magnitude for non-magnetic nanobuds, while for magnetic nanobuds the gaps have some finite difference between ↑ and ↓ eigen values. The $E_{gap}$ calculated by using both PBE and BLYP functional depicts almost similar behavior for carbon based systems. However, the electronic and magnetic properties of any system is dependent on the type of exchange-correlation functional used.

$C_{28}$ fullerene is magnetic and possesses a TMM of 4.0$\mu_B$. When it is bonded to ACNT or ZCNT, the resultant nanobud remains magnetic in its most stable state but their TMM decrease to 1.7$\mu_B$ and 2.0$\mu_B$ respectively. The [1+1] atom cycloaddition mode of $C_{28}$-(5,5) and $C_{28}$-(5,0) acquire highest MM of 2.6$\mu_B$ and 2.8$\mu_B$ respectively w.r.t. all other cycloadditions. To understand the origin of magnetism, the local MM shows that the connecting bond atoms contribute ~ 0.2% to 8% of TMM, whereas the second and third NN contributes majorly towards TMM (23% and 21%) for $C_{28}$-(5,5) nanobud, 34% and 20% for $C_{28}$-(5,0) nanobud respectively. The interactions between $C_{28}$ and CNTs are ferromagnetic as shown in spin density maps (Fig. 5(b-c)). The $E_{gap}$ for ↑ and ↓ states are equal for non-magnetic state except [2+2] $A_5$-$A_5$/P of $C_{28}$-(5,0), which is magnetic (Table 4) having equal ↑ and ↓ values. Due to unequal distribution of ↑ and ↓ electrons in $C_{28}$-CNB cycloadditions result the magnetism among these nanobuds.

Similar to $C_{28}$ cage, isolated $C_{36}$ is also magnetic having TMM of 2.0$\mu_B$. The combination of $C_{36}$ with ACNT or ZCNT result in magnetic nanobud with TMM of 1.92$\mu_B$ and 1.95$\mu_B$ respectively for most stable configurations. The [2+2] $A_6$-$A_6$/S, [2+2] $A_6$-$A_6$/P configurations of $C_{36}$-(5,5) and $C_{36}$-(5,0) have maximum TMM of 3.92$\mu_B$ and 4.0$\mu_B$. For $C_{36}$-(5,5) in [2+2] $A_6$-$A_6$/S, first and third NN carbon atoms contribute 30% and 26%, whereas $C_{36}$-(5,0) in [2+2] $A_6$-$A_6$/P the contribution of first and third NN carbon atoms to TMM is 28% and 37%. However, for both nanobuds, the connecting bond atoms contribute only 1% and 5% towards TMM. The spin density maps show ferromagnetic interactions in $C_{36}$-CNBs as ↑ electron density is higher than ↓ electron density (Fig. 5(d-e)). All the possible cycloadditions have significant magnetic order in HOMO-LUMO gaps of ↑ and ↓ states showing their magnetic behavior.

The combination of non-magnetic $C_{40}$ with either magnetic ACNT or non-magnetic ZCNT results in non-magnetic nanobud in their most stable configuration. When $C_{40}$ is bonded to ACNT in [2+2] $A_5$-$A_5$/T mode of configuration, the nanobud has highest TMM of 1.55$\mu_B$ w.r.t. other configuration of $C_{40}$-(5,5). The major contribution of 34% and 22% towards TMM is shown by first and second NN carbon atoms of $C_{40}$-(5,5) in [2+2] $A_5$-$A_5$/T. For $C_{40}$-(5,0), the highest TMM (1.56$\mu_B$) is shown when connect through [6+6] ring cycloaddition. The local MM shows that carbon atoms of first, second and third NN contribute ~1.6% to 3.0% towards TMM, whereas the remaining contribution is shown by CNT. The Spin density plots show that the interactions are ferromagnetic in nature for $C_{40}$-(5,5) in [2+2] $A_5$-$A_5$/T configuration (Fig. 5(f)). The non-magnetic $C_{40}$-CNBs have equal $E_{gaps}$ for ↑ and ↓ states but for magnetic nanobuds ↑ and ↓ states have some finite energy difference.

Therefore, the result suggests that all the nanobud configurations have shown significant variation in the magnetic moment w.r.t the type of cycloaddition. The origin of magnetization may be explained in term of structural distortion of fullerene and CNTs in nanobud cycloadditions, redistribution of charges from connecting carbon atoms and polarization of charges due to formation of connecting bonds. The MM is distributed unevenly, but the major contribution of MM arises on carbon atoms which are at a distance from the connecting bonds atoms due to redistribution of charges in ↑ and ↓ electron states. The contribution from connecting bond atoms towards TMM is very small which may be due to sp$^3$ hybridization of completing their valency by make four σ-bonds with neighboring carbon atoms [48]. The Spin density plot (Fig. 5(b-f)) show there is a difference in spin up and spin down densities on various carbon atoms. The spin up density dominates w.r.t. spin down density which results in the magnetic behavior of these CNBs [48].

**Table 3; Total Magnetic moment(TMM) and HOMO-LUMO gaps($E_{gap}$) for $C_n$-(5,5).**

| Carbon nanobuds $C_n$-(5,5) | Nanobud Cycloaddition Configurations | TMM ($\mu_B$) | $E_{gap}$ (eV) | | | |
|---|---|---|---|---|---|---|
| | | | ↑ (PBE functional) | ↓ | ↑ (BLYP functional) | ↓ |
| $C_{20}$-(5,5) | [2+2] $A_5$-$A_5$/S | 0.00 | 0.43 | 0.43 | 0.43 | 0.43 |
| | [1+1] | 0.18 | 0.27 | 0.37 | 0.24 | 0.31 |
| $C_{28}$-(5,5) | [2+2] $A_5$-$A_5$/S | 1.70 | 0.41 | 0.06 | 0.38 | 0.06 |
| | [1+1] | 2.60 | 0.04 | 0.08 | 0.04 | 0.13 |
| $C_{36}$-(5,5) | [2+2] $A_5$-$A_5$/S | 1.92 | 0.42 | 0.16 | 0.42 | 0.17 |
| | [2+2] $A_6$-$A_6$/S | 3.92 | 0.41 | 0.26 | 0.41 | 0.26 |
| $C_{40}$-(5,5) | [2+2] $A_6$-$A_5$/S | 0.00 | 0.24 | 0.24 | 0.24 | 0.24 |
| | [2+2] $A_5$-$A_5$/T | 1.55 | 0.31 | 0.19 | 0.32 | 0.19 |

**Table 4; Total Magnetic moment(TMM) and HOMO-LUMO gaps($E_{gap}$) for $C_n$-(5,0)**

| Carbon nanobuds $C_n$-(5,0) | Nanobud Cycloaddition Configurations | TMM ($\mu_B$) | $E_{gap}$ (eV) | | | |
|---|---|---|---|---|---|---|
| | | | ↑ (PBE functional) | ↓ | ↑ (BLYP functional) | ↓ |
| $C_{20}$-(5,0) | [2+2] $A_5$-$A_5$/AA | 0.00 | 0.31 | 0.31 | 0.29 | 0.29 |
| | [1+1] | 0.28 | 0.07 | 0.22 | 0.07 | 0.28 |
| $C_{28}$-(5,0) | [2+2] $A_5$-$A_5$/P | 2.00 | 0.29 | 0.29 | 0.32 | 0.32 |
| | [1+1] | 2.80 | 0.16 | 0.31 | 0.16 | 0.30 |
| $C_{36}$-(5,0) | [2+2] $A_5$-$A_5$/P | 1.95 | 0.24 | 0.09 | 0.26 | 0.09 |
| | [2+2] $A_6$-$A_6$/P | 4.00 | 0.14 | 0.29 | 0.16 | 0.32 |
| $C_{40}$-(5,0) | [2+2] $A_6$-$A_5$/P | 0.00 | 0.25 | 0.25 | 0.29 | 0.30 |
| | [6+6] | 1.56 | 0.15 | 0.17 | 0.15 | 0.19 |

**Table 5; Net charge transfer (q) in most stable CNB configurations, Negative value means electron transfer from CNT to fullerene.**

| Carbon nanobuds $C_n$-CNBs | Type of CNT | Most stable cycloaddition of CNBs | q(e) |
|---|---|---|---|
| $C_{20}$-CNBs | Armchair | [2+2] $A_5$-$A_5$/S | -0.048 |
|  | Zigzag | [2+2] $A_5$-$A_5$/AA | +0.024 |
| $C_{28}$-CNBs | Armchair | [2+2] $A_5$-$A_5$/S | -0.248 |
|  | Zigzag | [2+2] $A_5$-$A_5$/P | -0.062 |
| $C_{36}$-CNBs | Armchair | [2+2] $A_5$-$A_5$/S | -0.151 |
|  | Zigzag | [2+2] $A_5$-$A_5$/P | +0.050 |
| $C_{40}$-CNBs | Armchair | [2+2] $A_6$-$A_5$/S | -0.085 |
|  | Zigzag | [2+2] $A_6$-$A_5$/P | -0.031 |

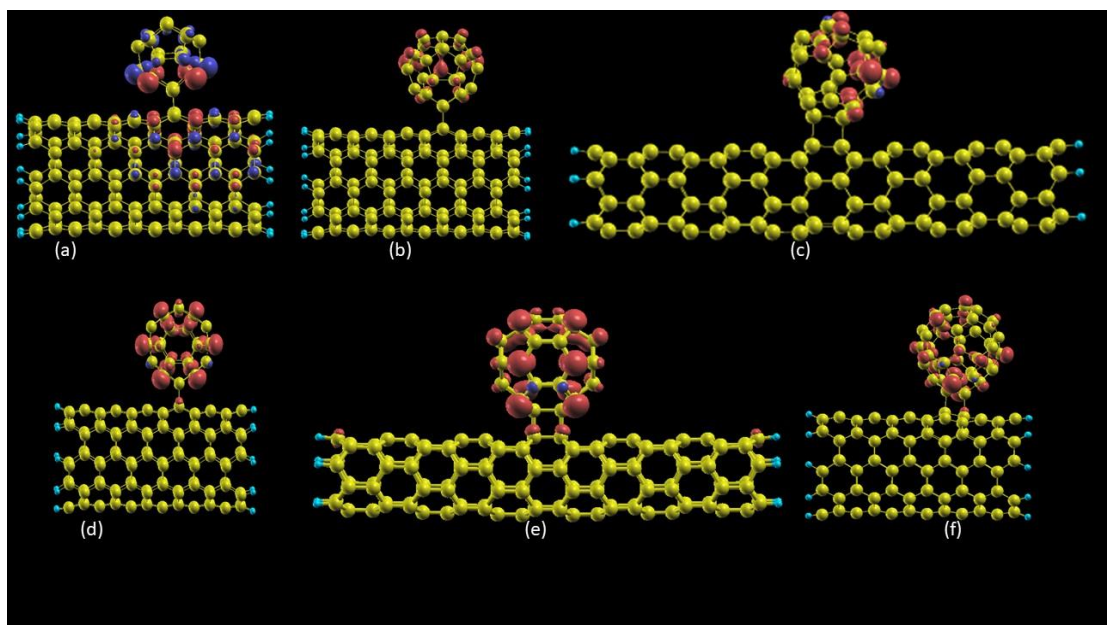

**Fig 5: - (in colour) Spin density maps; - (a) $C_{20}$-(5,5) [2+2] $A_5$-$A_5$/S (b) $C_{28}$-(5,5) [2+2] $A_5$-$A_5$/S (c) $C_{28}$-(5,0) [2+2]$A_5$-$A_5$/P (d) $C_{36}$-(5,5) [2+2]$A_5$-$A_5$/S (e) $C_{36}$-(5,0) [2+2]$A_6$-$A_6$/P (f) $C_{40}$-(5,5) [2+2]$A_5$-$A_5$/T.**

The Mulliken population charge analysis has been performed to check the change in net charge transfer as it provides a clear picture of charge redistribution in the system under investigation. This redistribution of charges at different carbon site is responsible for the variation in localized MMs. The attachment of fullerenes on CNTs changes the position of Fermi level of CNTs, which specifies the charge transfer between them (Table 5). For most stable nanobuds with ACNT base, the electron transfer takes place from CNT to fullerene but for ZCNT base, electrons are transferred from fullerene to CNT except $C_{28}$-CNBs and $C_{40}$-CNBs which have tendency of transferring the charge from CNT to fullerene. The pattern followed by charge transfer in $C_{20}$-CNBs is similar to a previous study [40]. The change in the direction of charge transfer might be due to difference in work function (WF) of fullerenes and CNTs. When the electron transfers from fullerenes to CNTs then it is expected that the WF of fullerenes is more than CNTs or vice versa. Previous results show that WF of ultra-small diameter ZCNTs is more than that of ACNTs [59], whereas WF of larger diameter CNTs are very similar. This is therefore consistent with the direction of charge transfer from CNTs to fullerene with ACNT base and vice versa for ZCNT bases. This behavior implies that electronic properties of these nanobuds can be tuned.

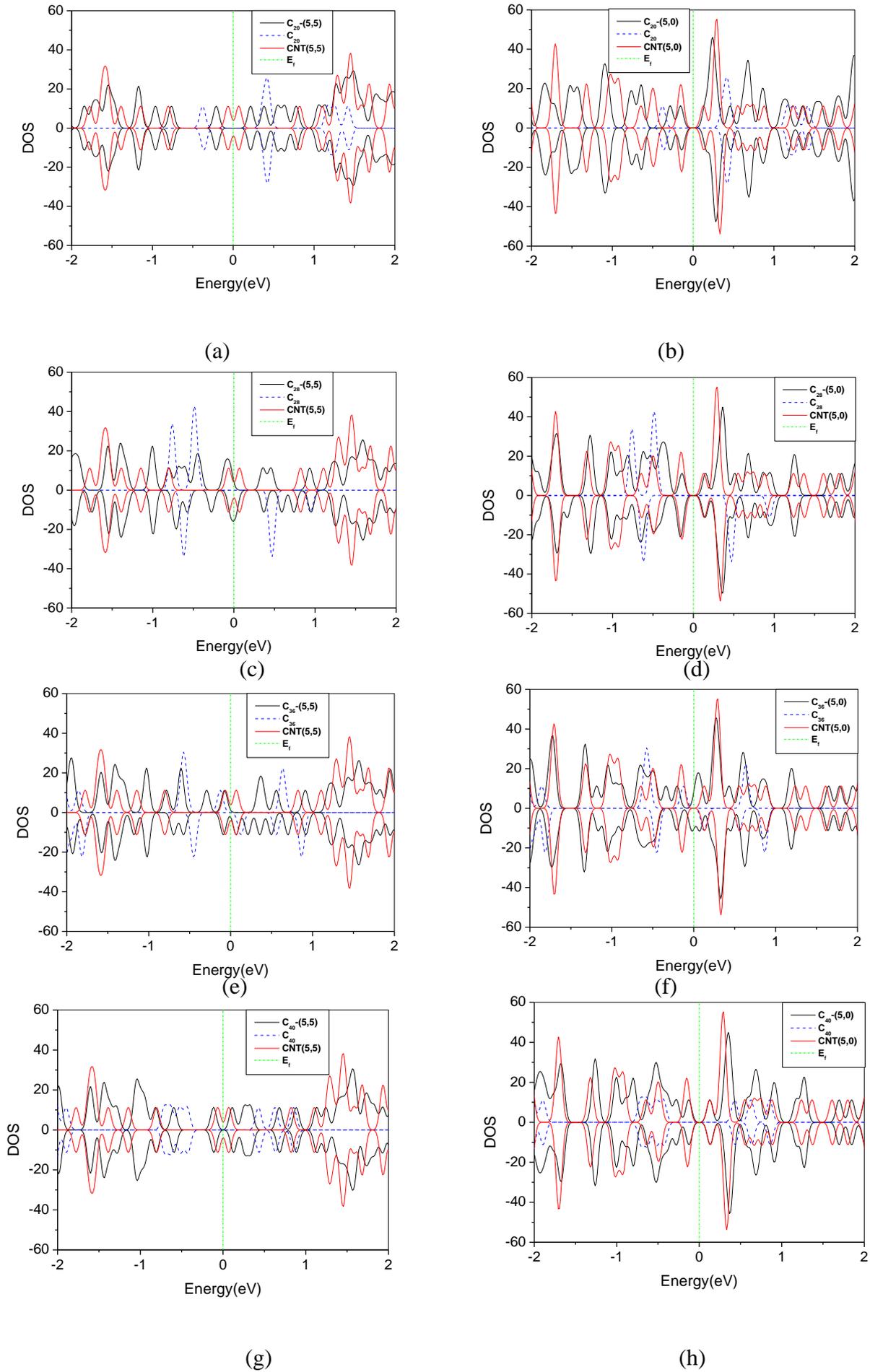

**Fig. 6; (in colour) Electron Density of States for the most stable CNBs.**

To further explore the change in electronic properties of CNT due to smaller fullerene ($C_{20}$, $C_{28}$, $C_{36}$, $C_{40}$) attachment, Electron Density of State (EDOS) for both ↑ and ↓ electrons are plotted for the most stable configurations of nanobuds (Fig.6). EDOS plots show that there is significant variation in spin up and spin down states near the fermi level after the formation of nanobuds. The ACNT is metallic and has non-zero density of states at the Fermi level whereas ZCNT is found to be semiconducting. Upon attachment of fullerenes ($C_{20}$ and $C_{40}$) on CNTs there is opening up of the gap that can be due to the effect of strong hybridization between the two systems which leads to semiconducting nanobud. The $C_{20}$-CNB results are in agreement with a previous reported DFT study [40]. The $C_{28}$-CNBs and $C_{36}$-CNBs with ACNT or ZCNT base show no opening of gap, but the attachment of these fullerenes introduced more impurity states near to Fermi level. The $C_{20}$-CNBs and $C_{40}$-CNBs show increase in gaps whereas the gap decreases for $C_{28}$-CNBs and $C_{36}$-(5,0). For $C_{36}$-(5,5), the gap remains almost constant. It can be predicted that there will be tuning of $E_{gaps}$ depending on the type of fullerenes attached to CNTs.

## **Conclusion**

We have investigated the structural, electronic and magnetic properties of armchair or zigzag CNBs using first-principle method where the smaller fullerenes $C_{20}$, $C_{28}$, $C_{36}$ and $C_{40}$ are covalently bonded to side wall of ACNT or ZCNT. The structural optimization of nanobuds show that the carbon atoms of CNT near the fullerene are pulled outwards from the original wall surface and their bonding thus get transformed from $sp^2$ to $sp^3$. Smaller fullerenes prefer to attach more with ZCNT than ACNT. The connecting C-C bond lengths lie within the range 1.56Å-1.61Å. The stabilities of these nanobuds depend upon the type of C-C bond in the cycloaddition reaction. The formation of these nanobuds is energetically favorable and they form most stable structures through [2+2] bond cycloaddition. The nature of nanobud formed depends on the mode of attachment of both subunits i.e. CNT and fullerene

The covalent bonding between fullerene and CNT result in significant change the electronic and magnetic properties w.r.t. isolated CNT. For the most stable nanobuds with ACNT base, the electrons are transferred from CNT to fullerene but for ZCNT base, electron can either be transferred from CNT to fullerene or vice-versa. The electron density of state plots reveals that for $C_{20}$-CNBs and $C_{40}$-CNBs, there is an opening of gap near Fermi level. However, there is an addition of impurity states near the Fermi level for $C_{28}$-CNBs and $C_{36}$-CNBs. Both $C_{28}$-CNBs and $C_{36}$ CNBs show semiconducting behavior with narrowing of HOMO-LUMO gap. In general, all the CNBs are semiconductor regardless of whether the original CNT base is metallic or semi conducting. When a non-magnetic fullerene ($C_{20}$ and $C_{40}$) is combined with either magnetic or non-magnetic CNT, the resultant nanobud is also found to be non-magnetic in its most stable position, whereas the combination of magnetic fullerene ($C_{28}$ and $C_{36}$) with both magnetic and non-magnetic CNT gives magnetic nanobud with finite value of magnetic moment. The magnetic moments are more localized on nearest neighbor carbon atoms rather than on the connecting bond atoms. With the change in type of fullerene attached to the CNT, the HOMO-LUMO gaps of CNBs can be tuned that leads to their significant magnetic properties which can be further be explored for electronic applications. However, keeping in view the limitations of DFT and scope of the level of DFT used in the present calculations, more accurate finite temperature DFT calculations shall be required to further confirm the electronic and magnetic behavior of these nanobuds, so as to facilitate them as good electronic as well as magnetic material having potential applications in field emission and spintronics.

## **Acknowledgements**

The authors are thankful to DST-SERB, Govt. of India for providing financial support and also to SIESTA group for providing their computational code.